\documentclass{article}
\usepackage{arxiv}
\usepackage[toc,page]{appendix}

\usepackage{cite}
\usepackage{amsmath,amssymb,amsfonts}
\usepackage{algorithm}
\usepackage{algorithmic} 
\usepackage{graphicx}
\usepackage{textcomp}
\usepackage{xcolor}

\usepackage[utf8]{inputenc}
\usepackage[T1]{fontenc}  
\usepackage{hyperref}      
\usepackage{url}            
\usepackage{booktabs}      
\usepackage{nicefrac}      
\usepackage{microtype}       
\usepackage{lipsum}		
\usepackage{doi}
\usepackage{multirow}
\usepackage{makecell}
\usepackage{rotating}

\usepackage{dblfloatfix}

\usepackage{natbib}

\DeclareMathOperator*{\argmax}{argmax}

\title{High-Dimensional Stock Portfolio Trading with Deep Reinforcement Learning}

\date{} 					

\author{{Uta Pigorsch}\\
	Schumpeter School of Business and Economics \\
	University of Wuppertal\\
	Wuppertal, Germany \\   
	\texttt{pigorsch@uni-wuppertal.de} \\
	\And
	Sebastian Schäfer \\
    Schumpeter School of Business and Economics \\
    University of Wuppertal \\
    Wuppertal, Germany\\
	\texttt{sebastian.schaefer@uni-wuppertal.de} \\
}

\date{}

\renewcommand{\headeright}{}
\renewcommand{\undertitle}{}
\renewcommand{\shorttitle}{High-Dimensional Stock Portfolio Trading with Deep Reinforcement Learning}

\hypersetup{
pdftitle={High-Dimensional Stock Portfolio Trading with Deep Reinforcement Learning},
pdfsubject={q-fin.PM, cs.LG, q-fin.CP},
pdfauthor={Uta ~Pigorsch, Sebastian ~Schäfer},
pdfkeywords={Deep Reinforcement Learning, Portfolio Management, Algorithmic Trading},
}

\begin{document}
\maketitle

\begin{abstract}
    This paper proposes a Deep Reinforcement Learning algorithm for financial portfolio trading based on Deep Q-learning \citep{mnih2013playing}. The algorithm is capable of trading high-dimensional portfolios from cross-sectional datasets of any size which may include data gaps and non-unique history lengths in the assets. We sequentially set up environments by sampling one asset for each environment while rewarding investments with the resulting asset's return and cash reservation with the average return of the set of assets. This enforces the agent to strategically assign capital to assets that it predicts to perform above-average. We apply our methodology in an out-of-sample analysis to 48 US stock portfolio setups, varying in the number of stocks from ten up to 500 stocks, in the selection criteria and in the level of transaction costs. The algorithm on average outperforms all considered passive and active benchmark investment strategies by a large margin using only one hyperparameter setup for all portfolios.
\end{abstract}


\section{Introduction}
Portfolio management includes the process of analyzing financial assets and estimating future returns and risks. As the amount of available data, especially for stocks, is rising, intelligent systems which automate or even completely control the portfolio management workflow can increase the investor's performance \citep{gu2020empirical}. Therefore, machine learning is widely used within portfolio management, e.g. for return predictions or risk evaluation as in \citet{patel2015predicting,fischer2018deep,wolff2020stock}. \par
Deep Reinforcement Learning (DRL) is the process of agents finding policies that are maximizing the cumulative rewards for a specific task resulting from learning through interaction with an environment. As a basis for its decision-making, the agent receives information about the state of the environment for every action it is supposed to take. The agent should become artificially intelligent by using non-linear function approximators like neural networks in order to improve its decision-making.
\par
In the literature exist mainly two approaches to apply DRL within the context of portfolio management. Training agents that are capable of trading a single asset which can include both long and short positions in the respective asset, and training agents to make capital assignments to a portfolio of assets, i.e. learning the optimal weighting policy. However, as we discuss in the following, these approaches suffer from limited flexibility and/or generalization capabilities. \par
Both methodologies come with substantial drawbacks. Training an agent on a single asset makes the environment, i.e. the sample size, relatively small. For that, considering a stock with a data history of ten years and daily datapoints gives only approx. 2530 samples whereas, e.g. state-of-the-art neural networks for image recognition are evaluated using datasets that are multiple orders of magnitude larger \citep{he2016deep}. Small datasets come with the drawback that noise-induced signals are fit and do not provide sufficient information for a generalizable solution to the problem. Moreover, out-of-sample results for single assets are rather weak in terms of their economic and statistical expressiveness. Finding a hyperparameter setup that performs reasonable well in-, as well as out-of-sample, can be found with tenable computational efforts. This makes solutions found for a single asset fairly restricted to the asset itself and are potentially not valid for other assets. These issues motivate to train agents that are capable of trading asset portfolios. However, when using neural networks to generate output for a vector of aset weights, a complete feature matrix for every period is required. Specifically, each period that contains a missing value cannot be used for training and might also make adjacent periods obsolete as periods are usually interconnected. This requires, e.g., a portfolio consisting of stocks listed for the same time period. As a consequence, the number of available assets to combine into a portfolio is rather limited. We make crucial extensions to Deep Q-learning \citep{mnih2013playing} to simultaneously benefit from large, cross-sectional datasets and portfolios of assets while still using environments with only one asset at a time for training. \par
Our methodology contains three major extensions. First, we sample one asset from a set of assets and set up a typical DRL environment for trading the single asset. Using asset-specific features in each state, the agent makes periodical decisions when to invest in the single asset. At each terminal period, the environment is reset with a newly sampled asset. Second, we reward investments in the single asset with its next period’s return. Reserving cash, i.e. not investing, is rewarded with the mean return of the set of assets in the next period. Lastly, when evaluating the agent, in every period, we build equally weighted portfolios with each asset the agent is willing to invest in. This methodology allows using large, cross-sectional datasets with stronger generalization due to the random sampling of assets while the reward function motivates the agent to actively avoid investments that the agent predicts to perform below average. \par

The remainder of the paper is structured as follows: Section 2 briefly introduces the concepts of DRL in the context of trading financial assets and provides a short review of the related literature. Our algorithm, which uses Deep Q-learning \citep{mnih2013playing}, is proposed in Section 3 and tested on US stock data in Section 4. Section 5 concludes and discusses implications for further studies.

\section{Background and related literature}
\subsection{Reinforcement Learning}
We assume that the task of trading a financial asset can be represented as a partially observable Markov decision problem (POMDP) \citep{spaan2012partially}. We do so as there must be some underlying state that describes every possible trading decision made in the markets but is as large that it is pratically unobservable. We therefore construct an observable subset of the state consisting of features that are sufficiently relevant. The features are outlined below. We follow the standard definitions for the POMDP and denote the environment by $\mathcal{E}$ in which the agent acts. For each iteration, the agent receives information about the state in time step $t$, denoted as $s_t$, and performs an action $a_t$ from the action space $\mathcal{A}$. As a consequence, the environment returns a reward $R_t$ from the reward function $\mathcal{R}$ given the specific state-action pair. Note that we use $R_t$ for rewards and $r_t$ for financial returns in the following. That is, for every iteration, the agent receives a tuple consisting of the state, action and reward, i.e. $(s_t, a_t, R_t)$. Based on the tuple, the agent's task is to find a profitable trading strategy, given by the policy $\pi(s_t)$.

\par

When trading financial assets, the agent must make periodical decisions. The definition of a period can hereby vary from split seconds to weeks. At each period, the agent receives information about the current state of the environment. In the context of stock trading specifically, this may be the agent's available capital and trading positions as well as asset-specific information such as figures from recent earnings reports and the price history. The latter is frequently represented by technical indicators. \par

Using the available information, the agent should decide on an action which results in a step in the environment. In a financial trading environment that means undertaking an investment or adjusting portfolio weights with the available capital. The reward is typically defined by the asset's or portfolio's return, either immediately for each period or cumulative at the end of multiple periods. Alternatively, the excess return compared to a specific benchmark or the Sharpe-Ratio can be well suited, too \citep{almahdi2017adaptive,yang2020deep,theate2021application}. \par

\subsection{Related work}
DRL is widely used for trading and portfolio management. These studies are either concerned with trading assets individually or trading portfolios by learning dynamic asset allocation policies. The following studies are concerned with the trading of single assets. \par
\citet{huang2018financial} demonstrate the effectiveness of DRL for trading currency pairs. The author uses recurrent DRL to trade 12 currency pairs and concludes that using a rather small replay memory, compared to what is used in standard reinforcement learning, can be more effective. Moreover, the author finds that higher trading costs do not necessarily decrease trading performance as learnt strategies become more robust. On all pairs, the agent successfully outperforms the benchmark strategies buy-and-hold as well as sell-and-hold. Moreover, the author provides insights into the trading behavior of the agents. That is, agents maintain higher win rates at approx. 60\% with average profits close to zero. These results are in line to our findings. Specifically, we show that the agents are capable of adjusting to increased transaction costs as compared to active benchmark strategies while also yielding higher cumulative returns than a passive investment approach. \par
The Deep Q-learning algorithm \citep{mnih2013playing} is a popular choice for automated financial asset trading. \citet{theate2021application} test trading strategies based on DQN-agents over a cross-section of 30 stocks and stock indices. The agents successfully find trading strategies that on average outperform a passive buy-and-hold strategy. The authors train a DQN agent for each asset individually over a period of six years and evaluate their respective performance on a subsequent two year time period. Overall, agents tend to alternate between a more passive, buy-and-hold strategy and a mean-reversion strategy but cannot outperform a passive investment benchmark for some assets. We find that the capability of alternating and combining multiple investment strategies to be crucial as we compare our agents to both passive and active bechmarks whereas none of these is a domninant strategy in the proposed portfolio setups. \citet{li2019application} compare different variants of the Deep Q-learning (vanilla vs. double vs. duelling) and find the best performance when using the vanilla Deep Q-learning on ten US stocks. Addtionally, \citet{zhang2020deep} compare different DRL-algorithms in discrete and continuous action spaces for trading futures contracts. The analysis contains 50 futures contracts from multiple asset classes. The authors compare a variety of DRL methodologies on each asset class by forming portfolios based on the decisions made on each asset. Overall, Deep Q-learning performs the best. Besides, the authors show robustness in the performance of the algorithms for various levels of transaction costs. Based on these results, we use the vanilla Deep Q-learning algorithm \citep{mnih2013playing} as our baseline algorithm. As our choice of the neural network architecture, we follow \citet{taghian2020learning} who compare different feature extraction neural network architectures for the task of DRL for financial asset trading. Based on tests on four assets, the authors find that, overall, a simple multi-layer-perceptron architecture based on a Deep Q-learning algorithm performs the best. Furthermore, they find the best results using the time series of raw price data (i.e. open, high, low, close prices) as inputs while comparing it with hand-crafted inputs from the time series of price data like candlestick patterns. As hand-crafted inputs seem to provide little predictive power, we only transform the time series of past returns using multiple moving averages in combination with stock specific data from quarterly statements to reduce the dimensionality of the input vector. \par   
Besides trading single assets, some studies address the portfolio allocation problem with DRL. \citet{xiong2018practical} show the effectiveness of DRL for portfolio trading using a DRL agent to trade a portfolio of 30 assets. The agent achieves an annualized return of 22.24\% in comparison to a 15.93\% return given by the Dow Jones Industrial. \citet{park2020intelligent} achieve promising results using deep Q-learning for multi-asset trading. The authors use an action mapping function to gain a discrete action space which is supposed to be more practical. The mapping function selects actions that are close in effect to the agent‘s chosen action while keeping the tradeover relatively low. With this setup, the authors achieve an outperformance over multiple benchmark strategies for a US and Korean portfolio. We follow the choice of a discrete action space and show the agent's capability to yield high cumulative returns even when transaction costs are high. \par
\citet{jiang2017deep} utilize an ensemble of parallelly trained agents to dynamically weight assets. Agents are trained independently and only share the last neural network‘s layer, i.e. the softmax. By applying the structure to the cryptocurrency market on a 30-minute time frame, the technique produces cumulative returns in multiples of the returns of the benchmark strategies and also strongly outperforms in terms of the Sharpe-Ratio. \citet{srivastava2020deep} use reinforcement learning to solve the problem of finding the optimal, dynamic asset allocation strategy. For that, the authors compare different network structures for the agents, specifically a convolutional neural network, vanilla recurrent neural network and a recurrent neural network with long short-term memory cells. Given a portfolio of 24 US stocks, all structures successfully outperform an equally weighted portfolio in terms of total returns and Sharpe-Ratio. Additionally, the authors present how feeding in the time series of past weights can stabilize trading and, thus, dramatically reduce the turnover chosen by the agents. \citet{almahdi2017adaptive} show that recurrent DRL agents perform well when optimizing for risk measures such as the Sharpe- or Calmer-Ratio. The authors validate their findings by comparing out-of-sample trading agents with several benchmarks, all being outperformed by the risk-measure-optimized agents. \par
Overall, these studies demonstrate the potential of DRL in portfolio management applications. We contribute to the existing literature by proposing a DRL algorithm that is highly flexible in the portfolio setup. That is, using our self-regularizing algorithm, one simple hyperparameter setup is sufficient to successfully trade a large variety of stock portfolios. As such, our approach is flexible, i.e. not tailored to individual stocks and, hence, more generally applicable. Furthermore, we completely disregard any hand-crafted, additional trading logic making the agent's trading strategies fully self-contained.

\section{Deep Q-learning for Portfolio Management}

A (risk-neutral) investor's goal is the maximization of wealth $W_T$, where $T$ is the terminal period at the end of an investment period, after undertaking a series of investment decisions that manipulate the initial wealth $W_{t=0}$. Starting with an initial wealth of $W_{t=0}$, the investor undertakes investment decisions like buying and selling stocks. In the following, one period corresponds to one day. That is, the agent can change its capital assignments daily. Hence, the investor's final wealth, $W_T$ is the cumulative return resulting from the daily decisions made by the agent, i.e.:
\begin{equation*}
    W_T = W_{t=0} \, \prod_{t=1}^T (1 + r_t),
\end{equation*}
with $r_t$ denoting the return at the end of day $t$.
To simplify the objective, we consider the natural logarithm of wealth, which is given by:
\begin{equation*}
    \log(W_T) = \log(W_{t=0}) + \sum_{t=1}^T \log(1 + r_t).
\end{equation*}
When the returns $r_t$ are close to zero, which is a reasonable assumption if the returns are measured on a daily frequency, we use the following approximation:
\begin{equation*}
    \sum_{t=1}^T \log(1 + r_t) \approx \sum_{t=1}^T r_t,
\end{equation*}
such that the objective becomes to maximize the sum of (daily) returns. However, as investors are myopic \citep{benartzi1995myopic}, short-term returns should be more valuable than returns in the distant future. Thus, we discount returns by a factor $\gamma$ such that the objective changes to maximizing $\sum_{t=1}^T \gamma^t r_t$. \par
As the agent must learn from its periodical rewards, a value should be assigned to each state-action pair. To this end, a state-action function is defined that assigns to each state-action pair a value based on the expected sum of future, discounted rewards assuming that the agent will follow the policy $\pi$ from the current timestep $k$ to $T$, i.e.:
\begin{equation}
    \begin{aligned}
        Q(s,a) & = \mathbb{E}_{\pi}\left[\sum_{k=0}^{T} \gamma^k R_{t + k} \, | \, s_t = s, a_t = a \right] \\
               & = \mathbb{E}_{a' \thicksim \pi}\left[R_t + Q(s', a') \, | \, s_t = s, a_t = a \right],
    \end{aligned}
    \label{eq:qFunction}
\end{equation}
where $\mathbb{E}_{\pi}$ is the expected value following policy $\pi$, $s'$ is the state that results from executing action $a$ in state $s$ and $a'$ is the action chosen from policy $\pi$ in state $s'$ (see also \citet{sutton2018reinforcement}).
A natural choice for a policy is to always select the action that maximizes the state-action value. However, this requires a model which determines, or, in a stochastic framework, approximates the state-action function $Q(s, a)$. \citet{mnih2013playing} suggest to solve this issue by training neural networks (here specifically called Q-networks) to approximate the state-action function. The authors use an agent that has a discrete action space and decides based on the state-action value predictions. The Q-network, in which parameters are randomly initiated, is trained iteratively using stochastic gradient descent. Using the squared loss between predicted state-action values and partially observed state-action values, the neural network is iteratively enhanced. A predicted state-action value is obtained from the Q-network with the state-vector as the input layer. Partially observed state-action values, $y_i$, are obtained based on \eqref{eq:qFunction} where the reward $R_t$ is observed and the state-action value is predicted by the Q-network, however, assuming that in the consecutive periods only optimal actions are chosen:
\begin{equation*}
    y_i = Q^*(s,a) = \mathbb{E} \big[R_t + \max_{a'} Q^*(s', a')
        | s_t = s, a_t = a \big].
\end{equation*}
If state $s$ is the terminal state then there are no more actions to take and no state-action value predictions necessary. Hence, the target is set to the reward that is observed for the state-action pair: $y_i = R_t$. Using the maximum expected state-action value makes the algorithm off-policy, i.e. the policy is updated using actions that are not originated by the policy itself. The squared loss, which should be minimized, results as:
\begin{equation*}
    L_i(\theta_i) = \mathbb{E}[(y_i - Q(s, a; \theta_i))^2],
\end{equation*}
with the corresponding gradient as:
\begin{equation} \label{eq:gradient}
    \begin{aligned} \nabla{\theta_i} L_i(\theta_i)
        = \mathbb{E}_{s'\thicksim \mathcal{E}} \big[ & (R_t + \gamma \max_{a'} Q (s', a'; \theta_{i-1})           \\
                                                     & - Q(s, a; \theta_i) \nabla{\theta_i} Q(s,a;\theta_i)\big].
    \end{aligned}
\end{equation}
Samples are generated via an epsilon-greedy strategy. That is, with a probability of $\epsilon$, the agent chooses a random action while it selects the greedy action, i.e. the action with the highest expected state-action value, with a probability of $1 - \epsilon$. This ensures that sufficient exploration is undertaken. However, the concrete value of $\epsilon$ depends on the problem itself. Since consecutive samples can be highly correlated, the authors propose using an experience replay memory. While exploring the environment, new samples are stored in the experience replay memory. When updating the Q-network's parameters via stochastic gradient descent, a batch of samples is randomly drawn from the experience replay memory. This furthermore increases the sample efficiency of the algorithm as samples are used more than once \citep{mnih2013playing}. The full algorithm is called Deep Q-learning. We use the algorithm as described and include further extensions such that it fits the purpose of trading financial asset portfolios. \par
Generally, the agent is supposed to be able to train portfolios of assets while it is also desirable that it can deal with different history lengths of the assets. To deal with that, we differentiate between a training and a trading phase. \par
When training the agent, we only construct environments with single assets. Here, the agent needs to learn to trade a single asset by either investing in the asset or holding cash. Our methodology is long-only which is a product of the reward function that is outlined further below. A single asset environment is constructed by randomly drawing one of the assets from the selection of available assets with replacement. That is, while training, each asset is shown to the agent possibly multiple times. We define the state as a stack of asset-specific features and of a dummy variable that indicates the most recent action the agent has made. Each time the agent reaches the terminal state of the asset's data history, a new asset is drawn, i.e. the environment is reset. \par

The process of optimal trading of financial portfolios by an agent can be defined as a maximization of wealth using a policy that assigns capital to a set of assets. We use a discrete action space as motivated by \citet{park2020intelligent}. The agent can choose between either $a_t = 1$ with which the agent invests in the asset or holds its long position if it already invested. If it decides for $a_t = 0$, the agent decides to reserve cash for other investments. That reduces the problem to the following decision made for each asset and period: Will the asset perform above-average or should the capital be reserved for other assets? To promote that decision making, we reward investments in the asset with the next period's return adjusted for transaction costs $C$ if they occur. The reward for reserving cash is the average next period's return of the set of assets. Thus, the reward function is defined as:
\begin{equation} \label{eq:rewardFunction}
    R_{t} = \begin{cases}
        r_{t+1, i} - (1 - a_{t-1})C               & \text{if } a_t = 1  \\
        \frac{1}{N_t} \sum_{j=1}^{N_t}{r_{t+1,j}} & \text{if } a_t = 0,
    \end{cases},
\end{equation}
where $r_{t+1, i}$ is the next period's return of the currently selected asset in the environment and $N_t$ is the number of stocks in the set of assets in period $t$. \par
\begin{algorithm}
    \caption{Deep Q-learning for portfolio trading}\label{alg:dqnma}
    \emph{Extensions to the original Deep Q-learning algorithm from \citep{mnih2013playing} are marked in italics.} \vspace{0.5mm}
    \begin{algorithmic}[1]
        \STATE Initialize experience replay memory $\mathcal{D}$ to capacity $\mathcal{N}$
        \STATE Initialize action-value function $Q$ with random weights
        \STATE \emph{Sample random asset and generate environment}
        \STATE \emph{Initialize highest cumulative return on validation set to $CR_v^* = 0$}
        \STATE \emph{Initialize steps without evaluation to $\omega = 0$}
        \FOR {$t=1$ to $N$}
        
        \STATE \emph{$\omega = \omega + 1$
            \IF{$\omega = \Omega$}
            \STATE Get cumulative return on validation set and store in $CR_v$
            \IF{$CR_v > CR_v^*$}
            \STATE Update Q-network parameters with current parameters: $\theta^*$ $\leftarrow$ $\theta$
            \STATE Set $CR_v^* \leftarrow CR_v$
            \ENDIF
            \STATE Set $\omega = 0$
            \ENDIF}
        
        \STATE Get current state $s_t$
        \STATE With probability $\epsilon$ select random action $a_t$
        \STATE else select greedy action: $a_t = \argmax_a \, Q(s, a)$
        \STATE \emph{Update trading position and receive reward $R_t$ according to \eqref{eq:rewardFunction}}
        \STATE Observe next state $s_{t+1}$
        \STATE Store transition $(s_t, a_t, r_t, s_{t + 1})$ in $\mathcal{D}$
        \STATE Sample batch of $(s_j, a_j, r_j, s_{j + 1})$ from $\mathcal{D}$
        \STATE Set \\$y_j =
        \begin{cases}
            R_j                                          & \text{if state is terminal} \\
            R_j + \gamma max_{a'} Q(s_{j+1}, a'; \theta) & \text{otherwise}
        \end{cases} $
            \STATE Perform gradient step on $(y_j - Q(s_j, a_j; \theta))^2$ according to \eqref{eq:gradient}
        \STATE \emph{\textbf{if} state is terminal \textbf{then}}
        \begin{ALC@g}
            \STATE \emph{Sample new random asset and generate environment}
        \end{ALC@g}
        \STATE \emph{\textbf{end if}}
        \ENDFOR
    \end{algorithmic}
\end{algorithm}

This training setup enforces the agent to maximize the cumulative return in the training set. However, it is desired to find a policy that achieves high cumulative returns in out-of-sample datasets. Thus, we prevent the agent from overfitting by using a validation set. We use this set in regular evaluation intervals $\Omega$ by computing the out-of-sample cumulative return achieved by the agent's policy. We initialize the cumulative return on the validation set with $CR_{v}^* = 0$. That is, there will be no solution if the agent yields losses when trading in the validation set. However, this does not occur in our experiments. For brevity, we define the steps without evaluation as $\omega$. We save the Q-network's parameters with which the agent performs best on the validation set in terms of cumulative return. The full algorithm is stated in Algorithm \ref{alg:dqnma} \par
To compute cumulative portfolio returns, we compute state-action values for the assets in each period for both possible actions, i.e. investing and reserving cash. For that, we construct an equally-weighted portfolio from the assets that the agent wants to invest in, i.e. where the state-action values suggest a long position. As the agent can hold assets for multiple periods but the weight of the asset may still change, we account for the costs that occur when the weights need to be increased. These rebalancing costs occur when assets remain in the portfolio for multiple consecutive periods while the portfolio shrinks in the number of assets.

\section{Experiments}
\subsection{Methodology}
We apply our algorithm to a cross-section of US stocks. To this end, we select a data history of 500 stocks including daily data from 2010-01-01 to 2021-06-30. To evaluate the algorithm's capabilities we compare cumulative returns on the test period with three benchmarks. First, we use a passive approach via an equally weighted buy-and-hold portfolio using all stocks. Second, we include two active investment strategies following momentum or reversion with a simple rule. The momentum strategy invests in stocks with a positive average return over the last five trading days. In contrast, the reversion strategy buys stocks that have a negative average return over the last five trading days. We report cumulative returns for all strategies and considered sets of stocks. As the 500 stocks are selected based on their terminal market capitalization, the resulting returns are higher than those of stock indices, like the S\&P500. \par
We construct 15 portfolios from the cross-section. These differ in the number of stocks $k$ and the selection criteria. That is, we form portfolios with $k$ being 10, 25, 50, 100 and 200. For each $k$, we build portfolios using the biggest and smallest $k$ stocks in terms of market capitalization and one where $k$ stocks are randomly selected from the full cross-section. This gives each portfolio unique characteristics although they partly may share stocks. Besides, we also test the algorithm using all 500 stocks available. After selecting stocks for each portfolio, we drop for each stock those periods that contain a missing value. However, we keep the period’s data for all other stocks, as the algorithm is capable of trading a varying portfolio size and, hence, is not affected by data gaps. We split every portfolio's data history into a training, validation and test set. Validation starts on 2019-01-01, testing at 2020-01-01. This yields a test period that is a great challenge for the agent since it includes periods with sharp losses at the beginning of the Covid-19 pandemic coupled with a strong trend change resolving into a market with overall large gains while the agent is trained over a less volatile period. This situation allows to access the true capabilities of the agent to adjust to extreme scenarios, but using the beginning of 2020, the agent is also tested during a more tranquil period. \par
We use both fundamental data available from quarterly statements and the stock price history. As of fundamental data we use the following indicators: sales per share, gross margin, operating margin, net profit margin, return on equity, return on assets, current ratio, debt ratio, market capitalization, the book to market equity ratio and the latest close price which are overall closely related to key factors found in \citet{fama2015five}. A key feature for trading financial assets via DRL is the price history and is typically used in the input layer, e.g. in \citet{zhang2020deep,theate2021application, almahdi2017adaptive,jiang2017deep}. To reduce the dimensionality of the input vector, we transform the price history using moving averages of the return history, both arithmetically and exponentially weighted. We hereby use windows of 5, 10, 20, 50, 100 and 200 days. Furthermore, we use rolling standard deviations of returns over a history length of 5, 10, 20, 50, 100 days. Lastly, the Q-network's input vector contains a dummy variable indicating if the agent has already invested in the currently observed stock. This may be relevant for deciding if a trade is worth the transaction costs. Features are scaled to be normally distributed with a zero-mean and unit standard deviation using the training dataset. Afterwards, scaling is only applied to the validation and test dataset. \par
In our analyis, daily returns are used to calculate the reward for each period. We assume that the agent can enter the market for any amount of shares at the trading day's opening price and close all positions at the closing price. Besides, we analyze the performances for three levels of transaction costs: 1, 5, 10 basis points (bps). We hereby follow the literature that uses DRL for trading. E.g., \citet{srivastava2020deep} assume 5 bps for a US stock portfolio, \citet{theate2021application} assume transaction costs from 0 to 20 bps for single stocks and \citet{zhang2020deep} analyze the performance achieved for futures contracts from 1 up to 45 bps. \par

\subsection{Hyperparameters}

One of the main advantages of using reinforcement learning next to supervised learning for the portfolio management task is the ability of the agents to not only make predictions about the next period but also to place that prediction into the scope of a long-term planning process using discounted rewards. However, a financially reasonable risk-adjusted discount factor, e.g. the daily average rate of return of a market index, is typically very close to one. This means that the agent has to make long-term predictions about the distant future development of the stock. On the other hand, we expect more myopic predictions to be more reliable. Similarly, \citet{theate2021application} state that there is a trade-off between long-term orientation and increasing uncertainty in returns far into the future. Still, there are examples in which a discount factor close to one worked out well, e.g. in \citet{huang2018financial} or \citet{chen2021reinforcement}. According to our expectation, we found that when the discount factor is too close to 1, the uncertainty about future returns becomes too high and the policy breaks down to a constant prediction that chooses the same action in every state. The issue resolves at a discount factor of 0.9, a value that is also used in the Deep Q-learning approach of \citet{park2020intelligent}. The same reasoning applies to the exploration rate. We found that a higher exploration of 30\% enables the agent to regularly find improvements to its policy and avoids getting stuck in local minima. We found that 3,000,000 training steps are sufficient for every setup and optimal solutions are likely to be found much earlier. As in \citet{mnih2013playing} we keep the experience memory size at 10\% of the number of training steps, i.e. at 300,000.
\begin{figure*}[t]
    \centering
    \includegraphics[width=1\textwidth]{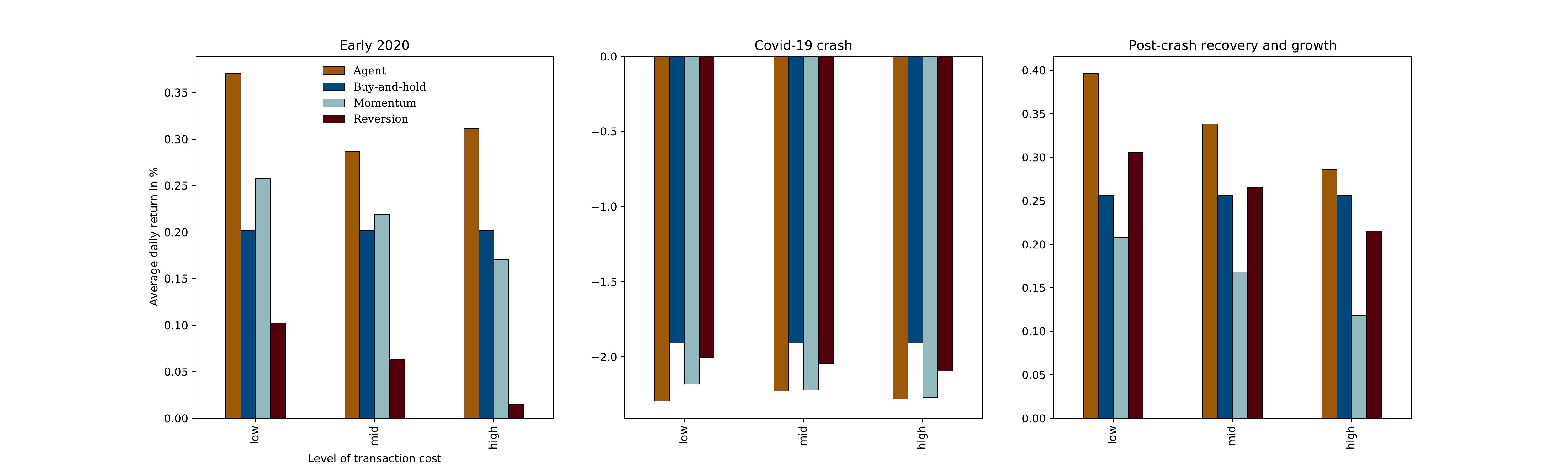}
    \caption{Out-of-sample trading performances for each level of transaction costs while splitting the test period into three phases based on market conditions.}\label{fig:PerformancePerMarketPhase}
\end{figure*}
Moreover, to increase the training efficiency, we do gradient steps every 20 iterations and use a large batch size of 1024 to efficiently use the capabilities of GPUs. Alongside, we evaluate the agent 300 times during training, i.e. every 10,000 iterations. It is important to note that the choice of the evaluation interval is associated with a trade-off between the additional time required for computations and performance that is lost due to unfrequent evaluations. \par
Financial data for return predictions contain a low signal-to-noise ratio. To avoid overfitting, we need to keep the number of parameters in the networks relatively low. We found it more useful to extend the number of layers instead of making the network wider, as, in much larger scales, is also done in \citet{he2016deep}. Hence, we use two hidden layers, activated by ReLU, with either 32, 64 or 128 neurons. Furthermore, we combine predictions from the three network configurations, forming an ensemble of predictors. \citet{krizhevsky2012imagenet} boost accuracy in an image classification task using an ensemble of convolutional neural networks. \citet{xie2013horizontal} show increased performance using ensembles of multiple networks for representation learning. Lastly, to update the weights of the agents' networks, we use the Adam optimizer, see also \citet{kingma2014adam}. All hyperparameters are summarized in Table \ref{table:Hyperparameters}. \par

\subsection{Results}

\begin{table}[t]
    \caption{List of hyperparameters used for all portfolio setups.}
    \begin{center}
        \begin{tabular}{ |c|c|c| }
            \hline
            Parameter               & Symbol        & Value     \\
            \hline
            Discount factor         & $\gamma$      & 0.9       \\
            Exploration probability & $\epsilon$    & 0.3       \\
            Number of iterations    & $N$           & 3,000,000 \\
            Experience memory size  & $\mathcal{N}$ & 300,000   \\
            Gradient step interval  & $-$           & 20        \\
            Evaluation interval     & $\Omega$      & 10,000    \\
            Optimizer               & $ - $         & Adam      \\
            Batch size              & $ - $         & 1024      \\
            Number of hidden layers & $ - $         & 2         \\
            \makecell{Number of neurons                         \\in hidden layers}              & $ - $         & 32 or 64 or 128        \\
            \hline
        \end{tabular}
    \end{center}
    \label{table:Hyperparameters}
\end{table}

For each of the portfolio and transaction costs setups, we train three agents (one for each Q-network setup) and form the respective ensembles. Then, we evaluate the out-of-sample performance using cumulative returns achieved in the test period compared to the three benchmark strategies. This gives in total 148 agents to train and 48 unique portfolio and transaction costs setups. All out-of-sample cumulative returns are reported in Table \ref{table:AllPerformances} (best performing strategies are indicated in bold) and outlined in greater detail in Fig. \ref{fig:Performances500} - \ref{fig:PerformancesHigh} in the appendices. Overall, the agent outperforms all benchmark strategies in 36 out of the 48 cases, and achieves the highest mean cumulative returns in every transaction costs setup. Individually compared to each of the benchmark strategies, the agent outperforms in 37, 44 and 44 out of the 48 cases the buy-and-hold, momentum, that buys stocks with recent positive returns, and reversion strategy, that buys stocks with recent negative returns, respectively. Agent outperformances are more consistent and higher when portfolios are larger, i.e. when more than 50 stocks are included. This is to be expected as the training dataset is growing only in the number of stocks and, thus, contains more information when the number of stocks is increased. Moreover, the agent shows the strongest relative outperformance on portfolios with smaller stocks. Both the agent and the active benchmark strategies (momentum and reversion) suffer from increasing transaction costs. However, the agent shows adaptiveness as the loss in performance is notably lower compared to the active benchmark strategies. Although the momentum strategy performs worse than the reversion strategy on average, there are setups in which the momentum strategy is superior to the reversion strategy. This indicates that the agents need to be highly dynamic in terms of the learnt strategies for each of the setups. \par
The test period can be split into three major phases around the crisis starting in March 2020. Before, the set of assets yields moderate, positive average returns in an uptrend followed by sharp losses during the crisis. Afterwards, the stock selection quickly recovers and realizes large gains. The agent's achieved mean returns in each of the three periods are compared to the benchmark strategies in Fig. \ref{fig:PerformancePerMarketPhase}. We find that the proposed algorithm performs especially well in the beginning of the test period and even improves the performance when increasing the transaction costs from 5 to 10 bps. However, the strategy does not outperform the benchmarks during the stock market losses in the beginning of the Covid-19 pandemic. As there is no major crisis in the training set, this result may be expected. In the case of recovering markets and larger gains, we find that the agent outperforms all benchmarks strategies on average which is in line with the results in early 2020.

\begin{table*}
    \caption{Out-of-sample cumulative returns at the end of the test period for each level of transaction costs and different portfolios}
    \centering
    \resizebox{0.67\textwidth}{!}{
        \begin{tabular}{|c|c|c|c|c|c|c|c|}
            \hline
            \textbf{Transaction costs} & \textbf{Portfolio size} & \textbf{Portfolio type} & \textbf{Agent}          & \textbf{Buy-and-hold} & \textbf{Momentum} & \textbf{Reversion} \\
            \hline
            \hline
            \multirow{17}{*}{1 bp}
                                       & \multirow{3}{*}{10}     & big                     & \textbf{222.9\%}        & 129.5\%               & 103.6\%           & 63.8\%             \\\cline{3-7}
                                       &                         & random                  & \textbf{68.1\%}         & 56.4\%                & 50.6\%            & 29.7\%             \\\cline{3-7}
                                       &                         & small                   & \textbf{109.0\%}        & 65.7\%                & 100.7\%           & 108.7\%            \\\cline{3-7}
            \cline{2-7}
                                       & \multirow{3}{*}{25}     & big                     & 41.4\%                  & \textbf{66.2\%}       & 39.4\%            & 58.8\%             \\\cline{3-7}
                                       &                         & random                  & \textbf{78.6\%}         & 48.1\%                & 47.6\%            & 38.2\%             \\\cline{3-7}
                                       &                         & small                   & \textbf{147.8\%}        & 84.0\%                & 108.5\%           & 94.1\%             \\\cline{3-7}
            \cline{2-7}
                                       & \multirow{3}{*}{50}     & big                     & \textbf{101.0\%}        & 52.7\%                & 44.5\%            & 60.9\%             \\\cline{3-7}
                                       &                         & random                  & \textbf{90.8\%}         & 62.5\%                & 16.9\%            & 79.9\%             \\\cline{3-7}
                                       &                         & small                   & \textbf{119.0\%}        & 66.9\%                & 51.7\%            & 85.9\%             \\\cline{3-7}
            \cline{2-7}
                                       & \multirow{3}{*}{100}    & big                     & \textbf{126.0\%}        & 56.3\%                & 53.8\%            & 90.1\%             \\\cline{3-7}
                                       &                         & random                  & \textbf{185.9\%}        & 67.8\%                & 17.1\%            & 116.6\%            \\\cline{3-7}
                                       &                         & small                   & \textbf{475.0\%}        & 68.0\%                & 52.2\%            & 72.0\%             \\\cline{3-7}
            \cline{2-7}
                                       & \multirow{3}{*}{200}    & big                     & \textbf{115.6\%}        & 58.9\%                & 31.0\%            & 111.5\%            \\\cline{3-7}
                                       &                         & random                  & 84.4\%                  & 63.6\%                & 17.7\%            & \textbf{106.1\%}   \\\cline{3-7}
                                       &                         & small                   & \textbf{195.0\%}        & 59.4\%                & 41.0\%            & 76.5\%             \\\cline{3-7}
            \cline{2-7}
                                       & 500                     & all                     & \textbf{155.1\%}        & 59.2\%                & 40.4\%            & 89.5\%             \\\cline{3-7}
            \cline{2-7}
                                       & \emph{Mean}             &                         & \emph{\textbf{144.7\%}} & \emph{66.6\%}         & \emph{51.1\%}     & \emph{80.1\%}      \\
            \hline
            \hline
            \multirow{17}{*}{5 bps}
                                       & \multirow{3}{*}{10}     & big                     & 123.2\%                 & \textbf{129.5\%}      & 76.6\%            & 44.3\%             \\\cline{3-7}
                                       &                         & random                  & 24.2\%                  & \textbf{56.4\%}       & 31.1\%            & 13.1\%             \\\cline{3-7}
                                       &                         & small                   & \textbf{173.7\%}        & 65.7\%                & 73.6\%            & 81.7\%             \\\cline{3-7}
            \cline{2-7}
                                       & \multirow{3}{*}{25}     & big                     & 11.6\%                  & \textbf{66.2\%}       & 20.4\%            & 37.1\%             \\\cline{3-7}
                                       &                         & random                  & 38.5\%                  & \textbf{48.1\%}       & 27.4\%            & 19.2\%             \\\cline{3-7}
                                       &                         & small                   & \textbf{134.6\%}        & 84.0\%                & 79.9\%            & 67.5\%             \\\cline{3-7}
            \cline{2-7}
                                       & \multirow{3}{*}{50}     & big                     & \textbf{76.0\%}         & 52.7\%                & 24.7\%            & 38.8\%             \\\cline{3-7}
                                       &                         & random                  & \textbf{79.2\%}         & 62.5\%                & 0.6\%             & 54.8\%             \\\cline{3-7}
                                       &                         & small                   & 64.3\%                  & \textbf{66.9\%}       & 30.7\%            & 60.2\%             \\\cline{3-7}
            \cline{2-7}
                                       & \multirow{3}{*}{100}    & big                     & \textbf{104.4\%}        & 56.3\%                & 32.5\%            & 63.6\%             \\\cline{3-7}
                                       &                         & random                  & \textbf{107.3\%}        & 67.8\%                & 0.8\%             & 86.5\%             \\\cline{3-7}
                                       &                         & small                   & 2\textbf{04.1\%}        & 68.0\%                & 30.9\%            & 48.0\%             \\\cline{3-7}
            \cline{2-7}
                                       & \multirow{3}{*}{200}    & big                     & \textbf{89.0\%}         & 58.9\%                & 12.8\%            & 82.0\%             \\\cline{3-7}
                                       &                         & random                  & \textbf{103.1\%}        & 63.6\%                & 1.3\%             & 77.4\%             \\\cline{3-7}
                                       &                         & small                   & \textbf{99.0\%}         & 59.4\%                & 21.3\%            & 51.8\%             \\\cline{3-7}
            \cline{2-7}
                                       & 500                     & all                     & \textbf{100.8\%}        & 59.2\%                & 20.8\%            & 63.1\%             \\\cline{3-7}
            \cline{2-7}
                                       & \emph{Mean}             &                         & \emph{\textbf{95.8\%}}  & \emph{66.6\%}         & \emph{30.3\%}     & \emph{55.6\%}      \\
            \hline
            \hline
            \multirow{17}{*}{10 bps}
                                       & \multirow{3}{*}{10}     & big                     & 127.6\%                 & \textbf{129.5\%}      & 47.8\%            & 23.3\%             \\\cline{3-7}
                                       &                         & random                  & 3.5\%                   & \textbf{56.4\%}       & 10.2\%            & -4.7\%             \\\cline{3-7}
                                       &                         & small                   & \textbf{68.3\%}         & 65.7\%                & 44.7\%            & 52.7\%             \\\cline{3-7}
            \cline{2-7}
                                       & \multirow{3}{*}{25}     & big                     & 46.1\%                  & \textbf{66.2\%}       & 0.2\%             & 14.2\%             \\\cline{3-7}
                                       &                         & random                  & -41.0\%                 & \textbf{48.1\%}       & 5.8\%             & -0.9\%             \\\cline{3-7}
                                       &                         & small                   & \textbf{114.3\%}        & 84.0\%                & 49.5\%            & 39.3\%             \\\cline{3-7}
            \cline{2-7}
                                       & \multirow{3}{*}{50}     & big                     & \textbf{53.8\%}         & 52.7\%                & 3.6\%             & 15.4\%             \\\cline{3-7}
                                       &                         & random                  & 55.2\%                  & \textbf{62.5\%}       & -16.6\%           & 28.4\%             \\\cline{3-7}
                                       &                         & small                   & \textbf{107.7\%}        & 66.9\%                & 8.5\%             & 32.9\%             \\\cline{3-7}
            \cline{2-7}
                                       & \multirow{3}{*}{100}    & big                     & \textbf{69.4\%}         & 56.3\%                & 9.9\%             & 35.7\%             \\\cline{3-7}
                                       &                         & random                  & \textbf{131.0\%}        & 67.8\%                & -16.4\%           & 54.7\%             \\\cline{3-7}
                                       &                         & small                   & \textbf{153.1\%}        & 68.0\%                & 8.5\%             & 22.6\%             \\\cline{3-7}
            \cline{2-7}
                                       & \multirow{3}{*}{200}    & big                     & \textbf{74.1\%}         & 58.9\%                & -6.5\%            & 50.8\%             \\\cline{3-7}
                                       &                         & random                  & \textbf{70.2\%}         & 63.6\%                & -16.0\%           & 47.0\%             \\\cline{3-7}
                                       &                         & small                   & \textbf{92.7\%}         & 59.4\%                & 0.5\%             & 25.8\%             \\\cline{3-7}
            \cline{2-7}
                                       & 500                     & all                     & \textbf{86.0\%}         & 59.2\%                & 0.1\%             & 35.1\%             \\\cline{3-7}
            \cline{2-7}
                                       & \emph{Mean}             &                         & \emph{\textbf{75.7\%}}  & \emph{66.6\%}         & \emph{8.4\%}      & \emph{29.5\%}      \\
            \hline
        \end{tabular}}
    \vspace{1mm}
    \label{table:AllPerformances}
\end{table*}

\section{Conclusion}
This paper proposes key extensions to Deep Q-learning \citep{mnih2013playing} to make the algorithm suitable for the trading of financial asset portfolios. We generate environments with single assets randomly drawn from a set of assets and put trading returns in the isolated environments into a portfolio-orientated perspective by rewarding cash reservation with the mean return of the set of assets. Furthermore, we use a validation set and reserve the best performing parameters in the Q-network to prevent overfitting. We use an ensemble of agents which further decrease overfitting and makes our methodology highly flexible in the number of assets. \par
With only a single hyperparameter setup, we test our methodology on 16 US stock portfolio configurations which vary in the number of stocks as well as the selection criteria. We benchmark the performance against a passive buy-and-hold and an active momentum and reversion strategy. Furthermore, we test three levels of transaction costs. The proposed methodology shows promising results, outperforming all benchmarks in 75\% of the setups. We find the algorithm to generate excess returns in rising market environments, however, it cannot avoid the sharp losses during the beginning of the pandemic. These findings motivate further extensions such as the inclusion of short-sales and accounting for risk, e.g. using the Sharpe-ratio.

\bibliographystyle{apalike}
\bibliography{references}

\begin{appendices}
    \section{Cumulative out-of-sample returns}
    The following figures show test returns for the respective portfolio setups achieved by the algotihm against the benchmarks.
    \begin{figure}[h!]
        \centering
        \includegraphics[width=0.9 \textwidth]{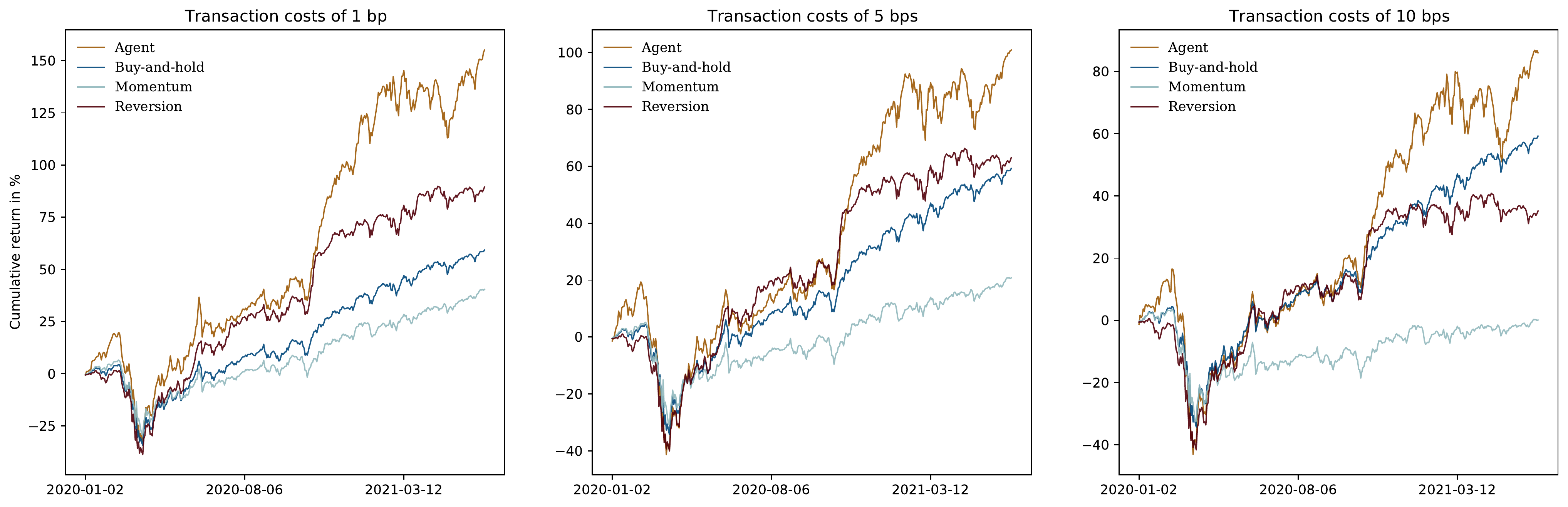}
        \caption{Test trading performances of the proposed agent compared to the benchmark strategies including all 500 stocks for the different levels of transaction costs.} \label{fig:Performances500}
    \end{figure}
    \begin{figure}
        \centering
        \includegraphics[width=0.9\textwidth]{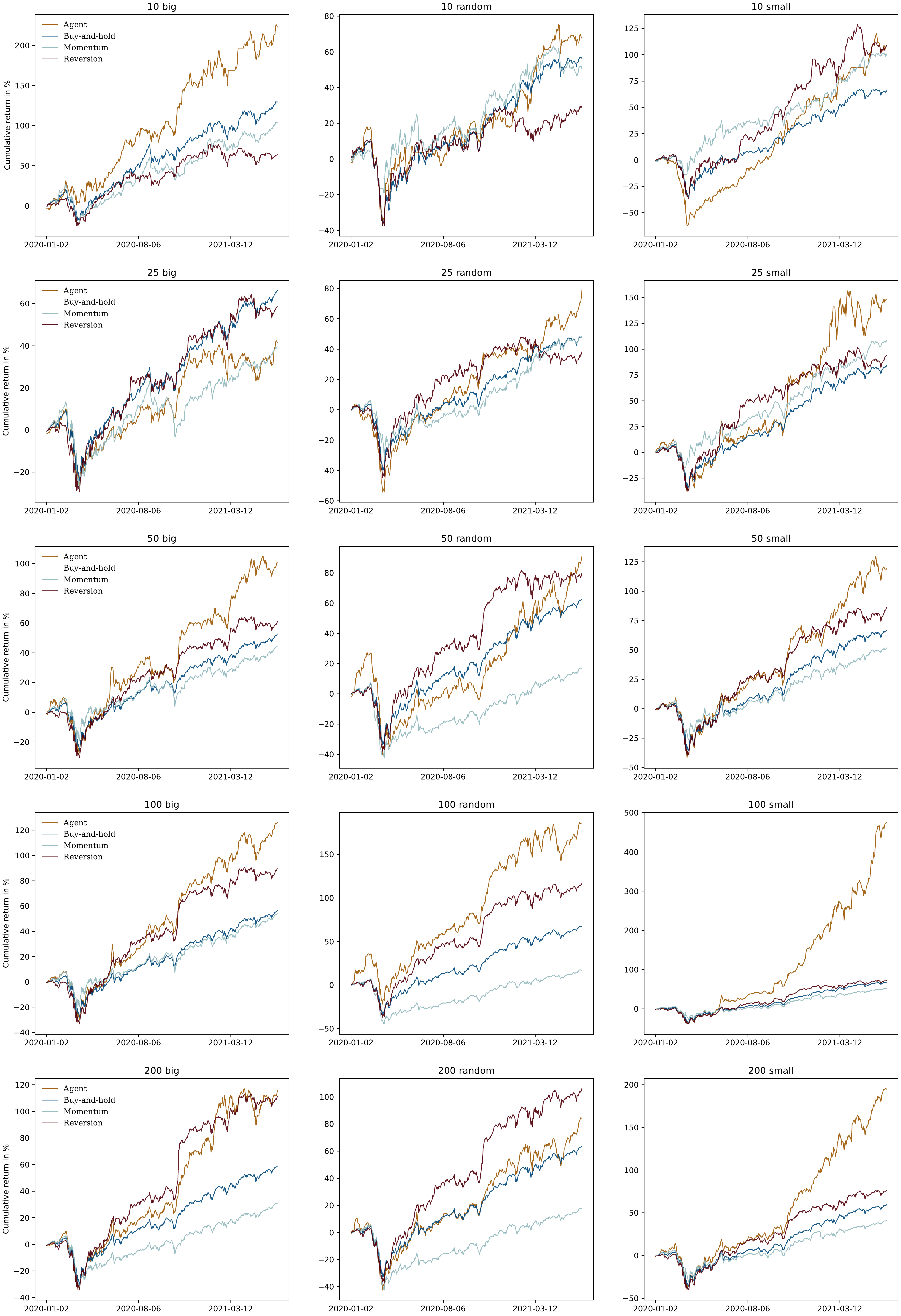}
        \caption{Test trading performances of the proposed agent compared to the benchmark strategies in the respective portfolio setup assuming transaction costs of 1 bp.}\label{fig:PerformancesLow}
    \end{figure}
    \begin{figure}
        \centering
        \includegraphics[width=0.9\textwidth]{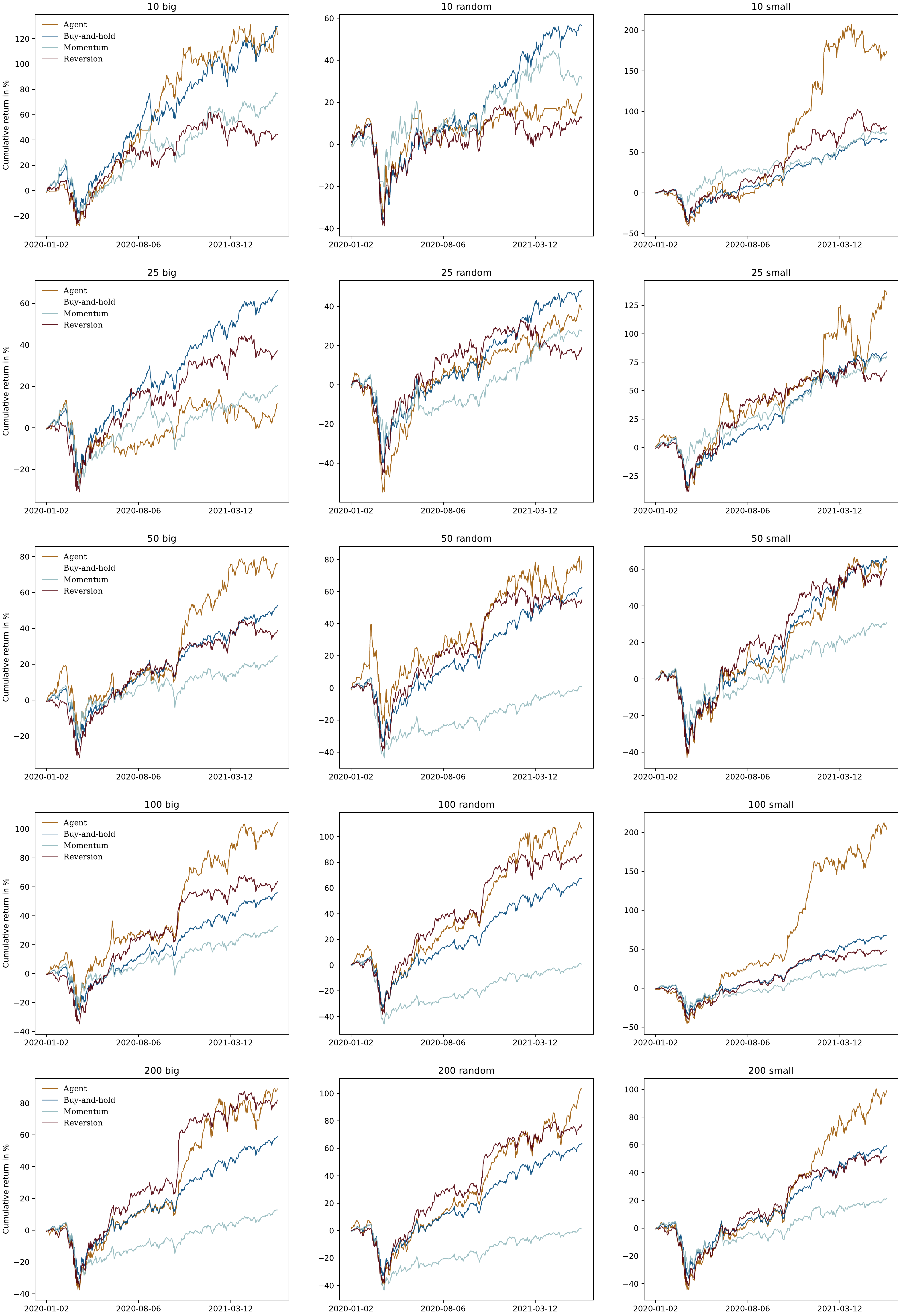}
        \caption{Test trading performances of the proposed agent compared to the benchmark strategies in the respective portfolio setup assuming transaction costs of 5 bps.}\label{fig:PerformancesMid}
    \end{figure}
    
    \begin{figure}
        \centering
        \includegraphics[width=0.9\textwidth]{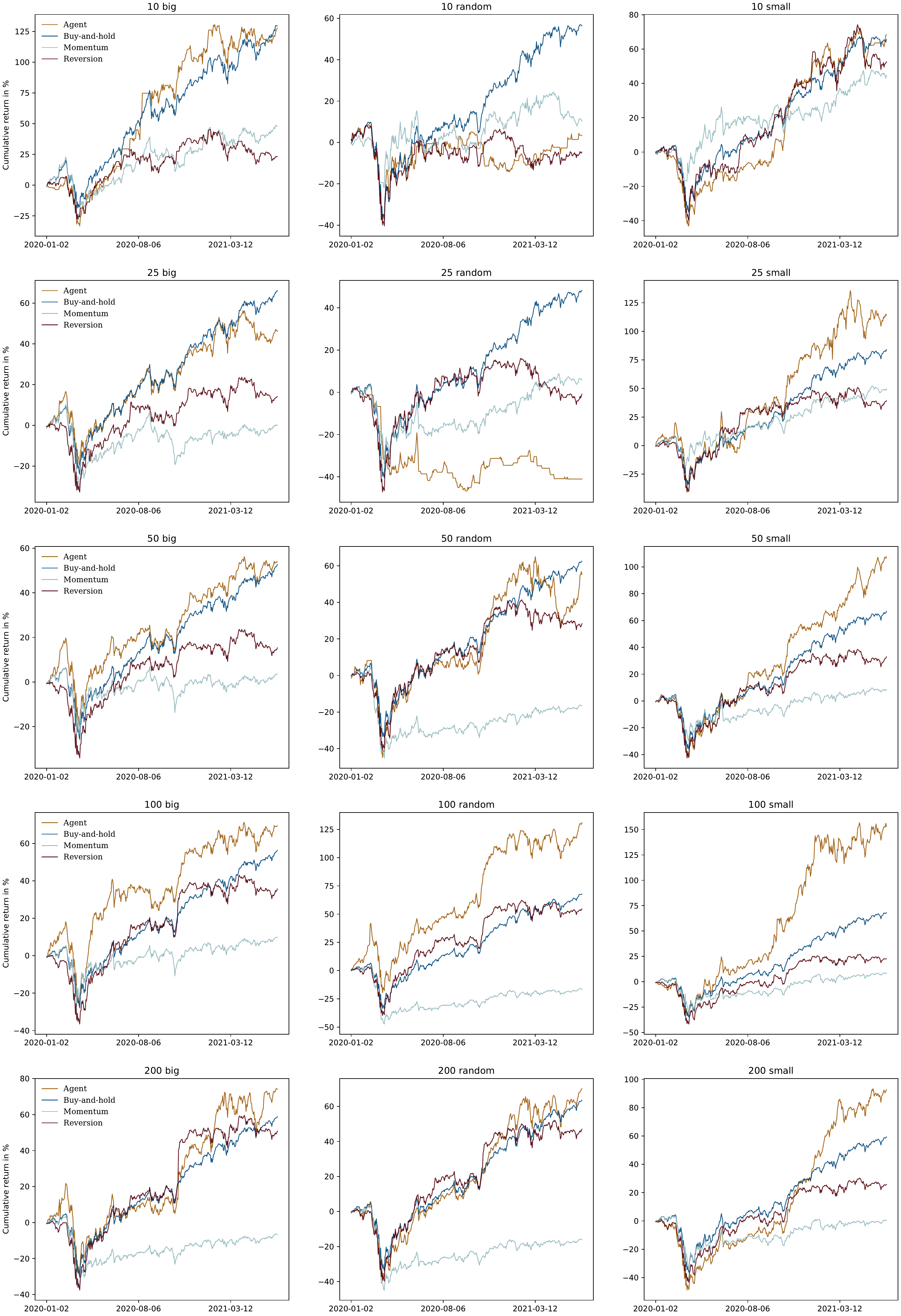}
        \caption{Test trading performances of the proposed agent compared to the benchmark strategies in the respective portfolio setup assuming transaction costs of 10 bps.}\label{fig:PerformancesHigh}
    \end{figure}

\end{appendices}

\end{document}